# A Wireless Sensor Network Air Pollution Monitoring System

Kavi K. Khedo[1], Rajiv Perseedoss[2] and Avinash Mungur[3]

Department of Computer Science and Engineering, University of Mauritius, Reduit, Mauritius
[1]k.khedo@uom.ac.mu
[2]arajivp@yahoo.com
[3]avi_29@hotmail.com

## ABSTRACT

*Sensor networks are currently an active research area mainly due to the potential of their applications. In this paper we investigate the use of Wireless Sensor Networks (WSN) for air pollution monitoring in Mauritius. With the fast growing industrial activities on the island, the problem of air pollution is becoming a major concern for the health of the population. We proposed an innovative system named Wireless Sensor Network Air Pollution Monitoring System (WAPMS) to monitor air pollution in Mauritius through the use of wireless sensors deployed in huge numbers around the island. The proposed system makes use of an Air Quality Index (AQI) which is presently not available in Mauritius. In order to improve the efficiency of WAPMS, we have designed and implemented a new data aggregation algorithm named Recursive Converging Quartiles (RCQ). The algorithm is used to merge data to eliminate duplicates, filter out invalid readings and summarise them into a simpler form which significantly reduce the amount of data to be transmitted to the sink and thus saving energy. For better power management we used a hierarchical routing protocol in WAPMS and caused the motes to sleep during idle time.*

## 1. INTRODUCTION

Sensor networks are dense wireless networks of small, low-cost sensors, which collect and disseminate environmental data. Wireless sensor networks facilitate monitoring and controlling of physical environments from remote locations with better accuracy [1]. They have applications in a variety of fields such as environmental monitoring, indoor climate control, surveillance, structural monitoring, medical diagnostics, disaster management, emergency response, ambient air monitoring and gathering sensing information in inhospitable locations [2, 3, 4, 5]. Sensor nodes have various energy and computational constraints because of their inexpensive nature and ad-hoc method of deployment. Considerable research has been focused at overcoming these deficiencies through more energy efficient routing, localization algorithms and system design.

In this paper we proposed a wireless sensor network air pollution monitoring system (WAPMS) for Mauritius. Indeed, with the increasing number of vehicles on our roads and rapid urbanization air pollution has considerably increased in the last decades in Mauritius. For the past thirty years the economic development of Mauritius has been based on industrial activities and the tourism industry. Hence, there has been the growth of industries and infrastructure works over the island. Industrial combustion processes and stone crushing plants had contributed to the deterioration of the quality of the air. Further, the economic success of Mauritius has led to a major increase in the number of vehicles on the roads, creating additional air pollution problem with smoke emission and other pollutants.

Air pollution monitoring is considered as a very complex task but nevertheless it is very important. Traditionally data loggers were used to collect data periodically and this was very time consuming and quite expensive. The use of WSN can make air pollution monitoring less complex and more instantaneous readings can be obtained [6, 7]. Currently, the Air Monitoring Unit in Mauritius lacks resources and makes use of bulky instruments. This reduces the flexibility of the system and makes it difficult to ensure proper control and monitoring. WAPMS will try to enhance this situation by being more flexible and timely. Moreover, accurate data with indexing capabilities will be able to obtain with WAPMS. The main requirements identified for WAMPS are as follows:

1. Develop an architecture to define nodes and their interaction
2. Collect air pollution readings from a region of interest
3. Collaboration among thousands of nodes to collect readings and transmit them to a gateway, all the while minimizing the amount of duplicates and invalid values
4. Use of appropriate data aggregation to reduce the power consumption during transmission of large amount of data between the thousands of nodes
5. Visualization of collected data from the WSN using statistical and user-friendly methods such as tables and line graphs
6. Provision of an index to categorize the various levels of air pollution, with associated colours to meaningfully represent the seriousness of air pollution
7. Generation of reports on a daily or monthly basis as well as real-time notifications during serious states of air pollution for use by appropriate authorities

At present, our scientific understanding of air pollution is not sufficient to be able to accurately predict air quality at all times throughout the country. This is where monitoring can be used to fill the gap in understanding. Monitoring provides raw measurements of air pollutant concentrations, which can then be analysed and interpreted. This information can then be applied in many ways. Analysis of monitoring data allows us to assess how bad air pollution is from day to day, which areas are worse than others and whether levels are rising or falling. We can see how pollutants interact with each other and how they relate to traffic levels or industrial activity. By analysing the relationship between meteorology and air quality, we can predict which weather conditions will give rise to pollution episodes.

## 2. RELATED WORKS

Wireless Sensor Network (WSN) is an active field of research due to its emerging importance in many applications including environment and habitat monitoring, health care applications, traffic control and military network systems [8]. With the recent breakthrough of Micro-Electro-Mechanical Systems (MEMS) technology [9] whereby sensors are becoming smaller and more versatile, WSN promises many new application areas in the near future. Typical applications of WSNs include monitoring, tracking and controlling. Some of the specific applications are habitat monitoring, object tracking, nuclear reactor controlling, fire detection, traffic monitoring, etc.

Initial development into WSN was mainly motivated by military applications. However, WSNs are now used in many civilian application areas for commercial and industrial use, including environment and habitat monitoring, healthcare applications, home automation, nuclear reactor controlling, fire detection and traffic control [8]. This transition from the use of WSN solely in military applications has been motivated due to the nature of WSNs which can be deployed in wilderness areas, where they would remain for many years, to monitor some environmental variables, without the need to recharge/replace their power supplies. Such characteristics help to overcome the difficulties and high costs involved in monitoring data using wired sensors. Below are some areas where WSN have been successfully deployed to monitor the environment.

## 2.1. Fire and Flood Detection

Large number of environmental applications makes use of WSNs. Sensor networks are deployed in forest to detect the origin of forest fires. Weather sensors are used in flood detection system to detect, predict and hence prevent floods. Sensor nodes are deployed in the environment for monitoring biodiversity.

The Forest-Fires Surveillance System (FFSS) [10] was developed to prevent forest fires in the South Korean Mountains and to have an early fire-alarm in real time. The system senses environment state such as temperature, humidity, smoke and determines forest-fires risk-level by formula. Early detection of heat is possible and this allows for the provision of an early alarm in real time when the forest-fire occurs, alerting people to extinguish forest-fires before it grows. Therefore, it saves the economical loss and environment damage. Similarly, a typical application of WSN for flood detection and prevention is the ALERT system [11] deployed in the US. Rainfalls, water level and weather sensors are used in this system to detect, predict and hence prevent floods. These sensors supply information to a centralized database system in a pre-defined way.

## 2.2. Biocomplexity Mapping and Precision Agriculture

Wireless sensor networks can be used to control the environment which involves monitoring air, soil and water. Sensors are deployed throughout the field and these sensors form a network that communicate with each other to finally reach some processing centre which analyse the data sent and then accordingly adjust the environment conditions (e.g., if the soil is too dry, the processing centre send signals which actuators recognise accordingly and thus can start the sprinkling system. Biocomplexity mapping system helps to control the external environment. Sensors are used to observe spatial complexity of dominant plant species [12]. An example is the surveillance of the marine ground floor where an understanding of its erosion processes is important for the construction of offshore wind farms [13].

Precision agriculture is an emerging WSN application area to monitor and control the amount of pesticides present in drinking water, monitor the level of soil erosion and the level of air pollution [14]. Precision agriculture encompasses different aspects such as monitoring soil, crop and climate in a field. Huge amount of sensor data from large-scale agricultural fields are frequently generated in such an application.

## 2.3. Habitat Monitoring

Concerns associated with the impacts of human presence in monitoring plants and animals in field conditions have to a large extent been overcome by WSNs [15]. Sensors can now be deployed prior to the onset of the breeding season and while plants are dormant or the ground is frozen as well as on small islets where it is unsafe or unwise to repeatedly attempt field studies. Such deployment represents a substantially more economical method for conducting studies than traditional personnel-rich methods where substantial proportion of logistics and infrastructure must be devoted to the maintenance of field studies, often at some discomfort and occasionally at some real risk.

Perhaps the best known application demonstrator for WSN in this domain is the Great Duck Island project at Berkley [15]. Sensors monitored the microclimates in and around nesting burrows used by the Leach's Storm Petrel in a non-intrusive and non-disruptive manner. Motes were deployed on the island, with each of them having a microcontroller, a low-power radio, memory, and batteries. Readings such as Infrared levels, humidity, rainfall and temperature were monitored on a constant basis to better understand the movements of the petrels. Motes periodically sampled and relayed their sensor readings to computer base stations on the island which in turn fed into a satellite link that allows researchers to access real-time environmental data over the Internet.

Researchers at University of Florida and University of Missouri, Colombia are studying the role of wildlife in maintaining diversity, tracking invasive species and the spread of emerging diseases by obtaining unobtrusive visual information. They are using DeerNet [16] which is a WSN-based system for analysing wildlife behaviour by tracking deer's actions. The overall goal is to develop a long-lived and unobtrusive wildlife video monitoring system capable of real-time video streaming. The captured video will be transmitted over to a remote monitoring center for real-time viewing and camera control. Advanced scene classification and object recognition algorithms together with fusion of data from other sensors like GPS and motion can be applied to remove essential visual information from the captured video. Then, statistical models about animals' food selection, activity patterns and close interactions can be made consequently.

## 3. RECURSIVE CONVERGING QUARTILES (RCQ) DATA AGGREGATION ALGORITHM

Most wireless sensor networks involve the collection of high amounts of data. For this reason, during last years considerable research effort has been devoted to data fusion and aggregation algorithms [17, 18]. In general, if we consider the problem to route data packets, representing measurements collected by sensors, to a single managing entity, i.e., a network sink, it is often efficient to exploit the correlation among similar data collected by the sensors in order to decrease overhead [19, 20]. At this point, however, a trade-off arises between the amount of transmitted data in the aggregated flows and their reliability. Data aggregation is a technique which tries to alleviate the localized congestion problem. It attempts to collect useful information from the sensors surrounding the event. It then transmits only the useful information to the end point thereby reducing congestion and its associated problems. We have developed a new data aggregation algorithm for WAPMS named Recursive Converging Quartiles (RCQ). The algorithm includes two basic operations namely duplicate elimination and data fusion.

### 3.1. Duplicate Elimination Technique

In WAPMS a packet consists of two parts: the data, which is the reading collected by the source node, and an id, which identifies the node uniquely in the network such as a network address. The cluster head collects readings from every node and stores them in a list. After collection, it goes through each item in the list and check for the occurrence of packets with the same id, thereby detecting the presence of duplicate packets. It then keeps only one instance of them. Figure 1and figure 2 illustrate our proposed duplication elimination technique.

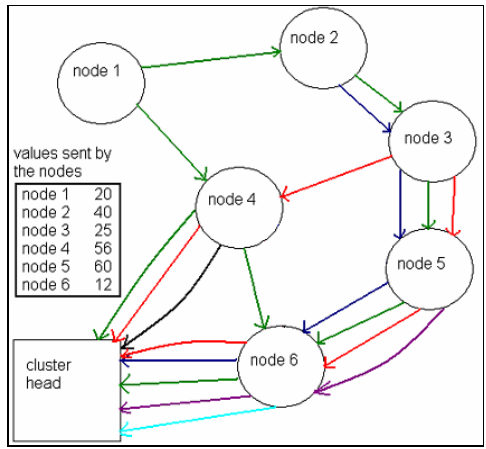

Figure 1. Multihop routing of data during a collection instance

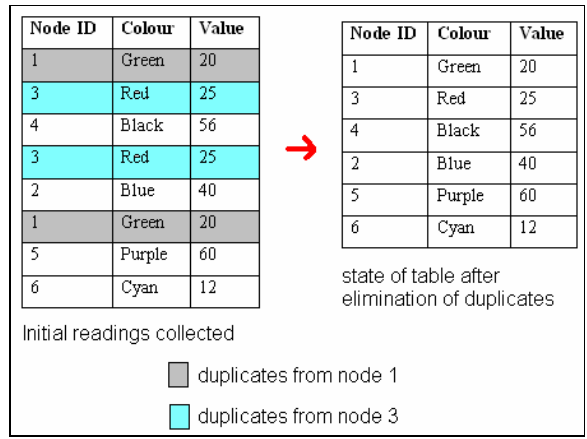

Figure 2. Illustration of duplicate elimination technique

### 3.2. Proposed Data Fusion Technique

There are several statistical methods to summarise a list of data. We have considered the use of the three quartiles - lower, median and upper. We have considered the use of quartiles since they are unaffected by extreme values; this is required in our system whereby extreme and invalid values can sometimes be transmitted to the cluster head and these should not influence the data fusion mechanism. Moreover, quartiles reduce the amount of data to only three values while still reflecting the original data in an accurate way. The novel data fusion algorithm works as follows:

1. The list is partitioned into several smaller groups
   - We consider the length of the list
   - We find its multiples in the form (x1, y1), (x2, y2)…
   - E.g., length = 200, multiples = (1, 200), (2, 100), (4, 50), (5, 40), (10, 20), (20, 10), (24, 5)
   - We choose the pair which will give the highest number of groups (Maximise x) and the minimum number of elements per group, while keeping it above a threshold (Minimise y, y > threshold value)

E.g., length = 50, multiples = (1, 50), (2, 25), (5, 10), (10, 5), threshold = 5, optimal pair = (10, 5).
2. We calculate the quartiles for each of the smaller lists
3. Merge the resulting quartiles for the sub lists into one list
4. Repeat the whole process until the eventual number of groups, in which the list can be broken, becomes one and the final list obtained has only three values.

Figure 3 below shows our proposed data fusion algorithm, Recursive Converging Quartiles, at work to achieve 3 values out of the original 33.

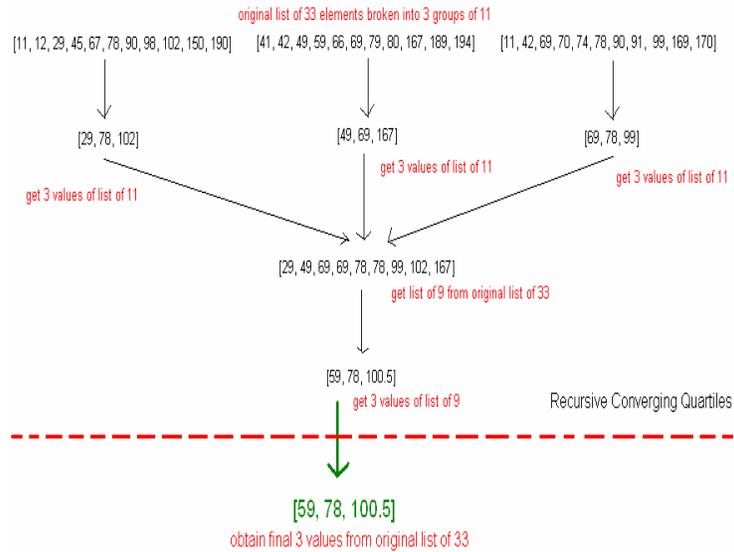

Figure 3. Using RCQ to aggregate a list of 33 values to only 3 values

## 4. WAPMS: THE PROPOSED AIR POLLUTION MONITORING SYSTEM

The proposed wireless sensor network air pollution monitoring system (WAPMS) comprises of an array of sensor nodes and a communications system which allows the data to reach a server. The sensor nodes gather data autonomously and the data network is used to pass data to one or more base stations, which forward it to a sensor network server. The system send commands to the nodes in order to fetch the data, and also allows the nodes to send data out autonomously. Figure 4 shows the architecture diagram of WAPMS.

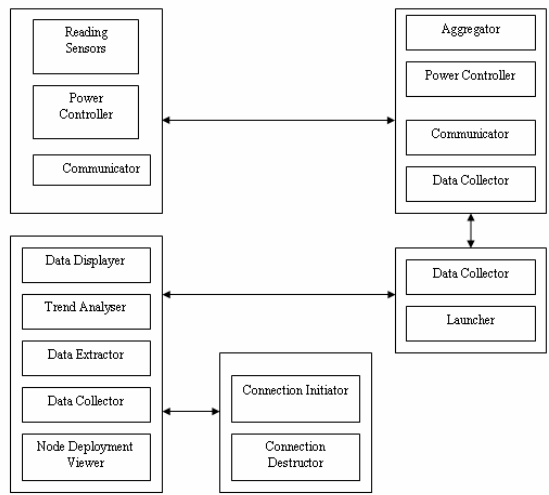

Figure 4. Architecture diagram of WAMPS

Below is a brief description of each component of WAMPS:

- *Reading Sensor:* generates a random value whose range is set based on the value of a "seriousness" variable.
- *Reading Transmitter:* gets the generated value from the reading sensor and transmits it through the communicator.
- *Power Controller:* Each node will have a method called "turn on" that will start the node and we just call it. As for power-saving modes, this will depend on what the simulator will provide to us.
- *Communicator:* this is implemented by the simulator. Inter-Process communication is usually done using sockets; so, we expect the simulator to provide us with sockets as well as methods such as "send" and "receive".
- *Launcher:* informs the data collector to start collection based on the delivery mode set by the user.
- *Data Collector:* gets a list of nodes from which it has to collect readings, then sends messages to inform them and finally receives the required values.
- Aggregator: implements the RCQ algorithm for data aggregation that we will discuss in the next section.
- *Data Extractor:* Use SQL queries to extract data from database
- *Data Displayer:* This extracts data as required by the user and displays them in a table as well as evaluates the AQI for the selected area.
- *Trend Analyser:* Gets previous readings and determines relationship between them to be able to extrapolate future readings.
- *Nodes Deployment Viewer:* Displays deployment of nodes in the WSN field and their AQI colours.
- *Connection Initiator:* The java DriverManager allows for a method to open a database, providing it the name of the database, user name and password as parameters. So, this component just has to make a call to this method and store the return reference to the connection.
- *Connection Destructor:* Connection object, in java.sql package, usually provides for a close method that closes the latter safely and frees associated memory as well as save the state of the latter. Therefore, this component just has to call this method.

The following table shows the various types of nodes that are present in WAPMS:

Table 1. Types of Nodes

| Type of Node | Energy Requirements | Location | Role |
|---|---|---|---|
| Source (sensor node) | Constrained | Random | Sensing and multihop routing |
| Cluster Head (collector) | Not-Constrained | Fixed | Collection and aggregation |
| Sink /Gateway | Not-Constrained | Fixed | Collection |

These nodes will form a hierarchy that is shown in figure 5 below:

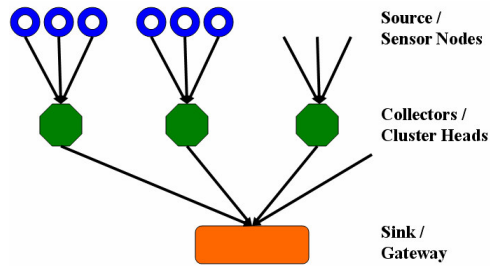

Figure 5. Hierarchy of nodes

The strategy to deploy the WSN for our system is as follows:

o We first partition our region of interest into several smaller areas for better management of huge amount of data that will be collected from the system and for better coordination of the various components involved
o We deploy one cluster head in each area; these will form cluster with the nodes in their respective areas, collect data from them, perform aggregation and send these back to the sink.
o We, then, randomly deploy the sensor nodes in the different areas. These will sense the data, send them to the cluster head in their respective area through multihop routing
o We will use multiple sinks that will collect aggregated from the cluster heads and transmit them to the gateway. Each sink will be allocated a set of cluster heads.
o The gateway will collect results from the sinks and relay them to the database and eventually to our application.

Figure 6 illustrates our deployment strategy:

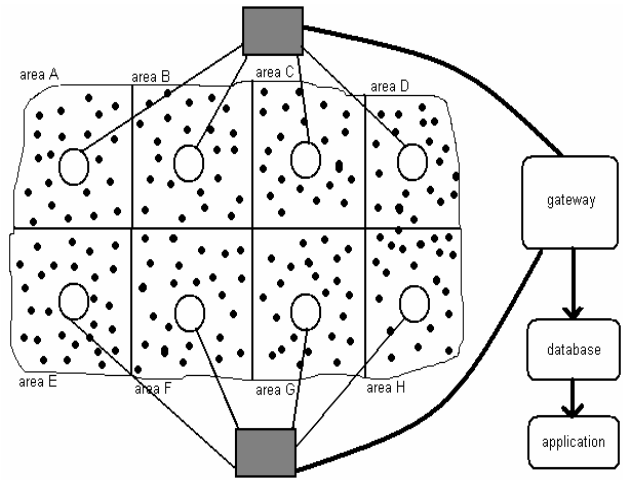

Figure 6. Deployment strategy

The system is simulated over a small region as a prototype and then it will be extended to the whole island. The town of Port Louis, the capital of the country, is chosen for the prototype implementation as it is an urban area and therefore, more exposed to air pollution than rural areas. The site is partitioned the site into 6 smaller areas as shown in figure 7. With this small number of areas, we will use a single sink and we further simplify the system by allowing the gateway to play the role of the latter.

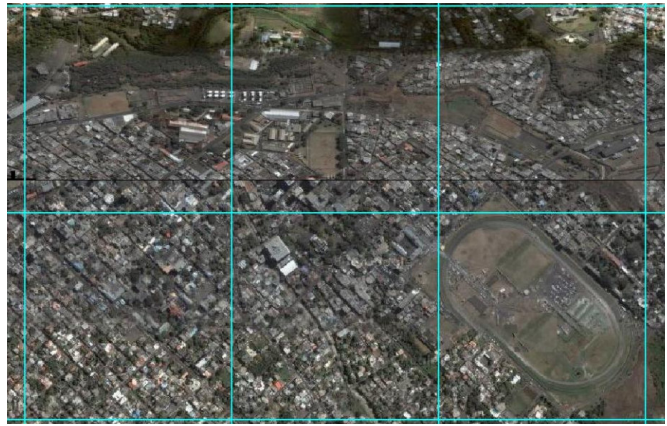

Figure 7. Partition of Port Louis into smaller areas

An Air Quality Index (AQI) is used in WAMPS. The AQI is an indicator of air quality, based on air pollutants that have adverse effects on human health and the environment. The pollutants are ozone, fine particulate matter, nitrogen dioxide, carbon monoxide, sulphur dioxide and total reduced sulphur compounds. Figure 8 and figure 9 illustrate the AQI range.

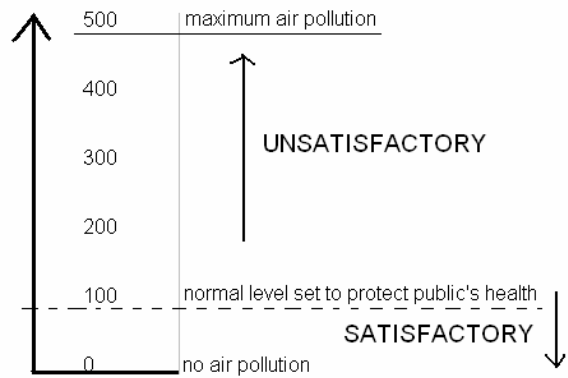

Figure 8. The range of AQI values

The AQI consists of 6 categories, each represented by a specific colour and indicating a certain level of health concern to the public and is it shown in figure 9. The Ambient Air Quality Standards for Mauritius reports that the safe limit for ozone is 100 micrograms per m$^3$ and the safe AQI value set is also 100. Therefore, the AQI itself can, indirectly, be used to measure ozone concentration in Mauritius.

| Air Quality Index Levels of Health Concern | Numerical Value | Meaning |
|---|---|---|
| Good | 0-50 | Air quality is considered satisfactory, and air pollution poses little or no risk. |
| Moderate | 51-100 | Air quality is acceptable; however, for some pollutants there may be a moderate health concern for a very small number of people who are unusually sensitive to air pollution. |
| Unhealthy for Sensitive Groups | 101-150 | Members of sensitive groups may experience health effects. The general public is not likely to be affected. |
| Unhealthy | 151-200 | Everyone may begin to experience health effects; members of sensitive groups may experience more serious health effects. |
| Very Unhealthy | 201-300 | Health alert: everyone may experience more serious health effects. |
| Hazardous | > 300 | Health warnings of emergency conditions. The entire population is more likely to be affected. |

Figure 9. Description of AQI categories

## 5. SIMULATIONS AND RESULTS

WAMPS has been simulated using the Jist/Swans simulator [21]. JiST is a high-performance discrete event simulation engine that runs over a standard Java virtual machine. It converts an existing virtual machine into a simulation platform, by embedding simulation time semantics at the byte-code level. SWANS is a scalable wireless network simulator built atop the JiST platform. SWANS is organized as independent software components that can be composed to form complete wireless network or sensor network configurations. Its capabilities are similar to ns2 and GloMoSim but it is able to simulate much larger networks. SWANS leverages the JiST design to achieve high simulation throughput, save memory, and run standard Java network applications over simulated networks. In addition, SWANS implements a data structure, called hierarchical binning, for efficient computation of signal propagation.

The DSR protocol [22] has been used for data transmission in WAPMS. The Dynamic Source Routing protocol is a simple reactive routing protocol designed specifically for use in multi-hop wireless ad hoc networks. DSR allows the network to be completely self-organizing and self-configuring, without the need for any existing network infrastructure or administration. DSR contains two phases: Route Discovery (find a path) and Route Maintenance (maintain a path). These only respond on a request. The protocol operates entirely on-demand, allowing the routing packet overhead of DSR to scale automatically to only that needed to react to changes in the routes currently in use.

After a collection, the system displays the nodes in their corresponding AQI colour as shown in figure 10. The following is an example of such a screen:

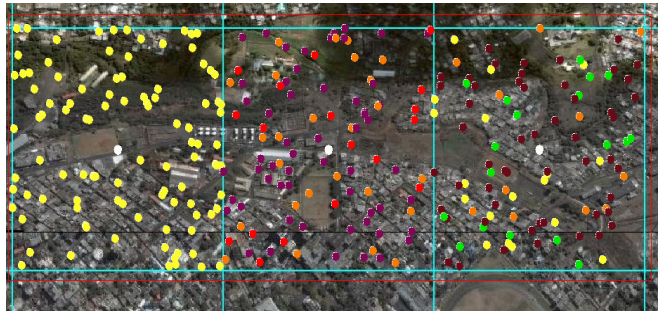

Figure 10. Nodes' Deployment after a collection

Given an area and a date, the system display the corresponding AQI readings and the health concern in this area as shown in figure 11 below.

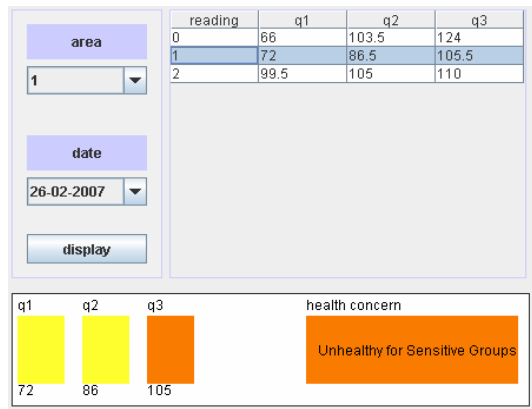

Figure 11. AQI for a selected area

Furthermore, the WAPMS system allows fast analysis of received data through line graphs of selected areas as shown in figure 12.

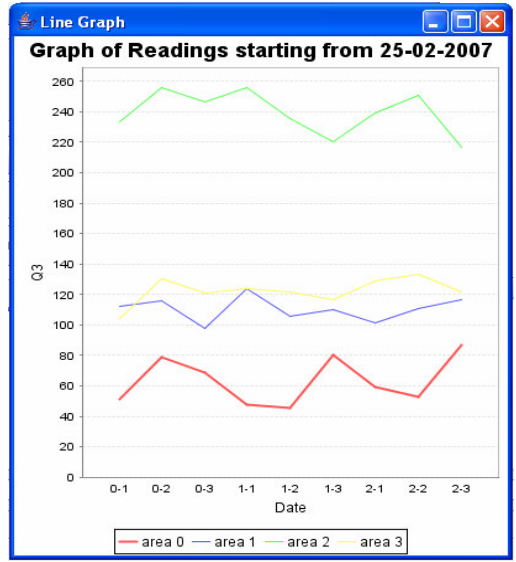

Figure 12. Line graph generated for selected areas

The performance of WAPMS has been evaluation with increasing load. We have varied the number of areas simulated from 1 to 6 and for each case, we have varied the number of nodes per area from 50 to 200 and the execution time of the system has been recorded. The results are shown in table II and figure 13.

Table 2. Execution Time of WAPMS

| No. of areas | Number of nodes per area | | | |
|---|---|---|---|---|
| | 50 | 100 | 150 | 200 |
| 1 | 20 | 58 | 110 | 170 |
| 2 | 23 | 80 | 200 | 330 |
| 3 | 34 | 115 | 280 | 500 |
| 4 | 44 | 150 | 380 | 690 |
| 5 | 50 | 200 | 470 | 860 |
| 6 | 56 | 250 | 540 | 985 |

As shown in the above table, the maximum running time of our simulator is less than 20 minutes in the worst case of 6 areas and 200 nodes in each area. The short execution time of WAPMS is massively advantageous comparing to the existing air pollution monitoring unit of Mauritius that often takes days to measure pollution in an area. Moreover, WAPMS allows timely monitoring of an area and an abnormal situation can be detected almost immediately.

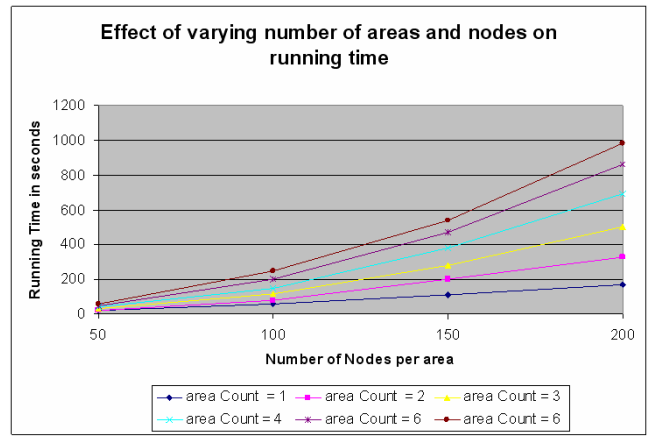

Figure 13. Performance analysis of WAPMS

## 6. CONCLUSION

As discussed in this paper, recent technological developments in the miniaturization of electronics and wireless communication technology have led to the emergence of Environmental Sensor Networks (ESN). These will greatly enhance monitoring of the natural environment and in some cases open up new techniques for taking measurements or allow previously impossible deployments of sensors. WAPMS is an example of such ESN. WAPMS will be very beneficial for monitoring different high risk regions of the country. It will provide real-time information about the level of air pollution in these regions, as well as provide alerts in cases of drastic change in quality of air. This information can then be used by the authorities to take prompt actions such as evacuating people or sending emergency response team.

WAPMS uses an Air Quality Index to categorise the various levels of air pollution. It also associates meaningful and very intuitive colours to the different categories, thus the state of air pollution can be communicated to the user very easily. The major motivation behind our study and the development of the system is to help the government to devise an indexing system to categorise air pollution in Mauritius. The system also uses the AQI to evaluate the level of health concern for a specific area.

WAPMS uses a novel technique to do data aggregation in order to tackle the challenge of power consumption minimisation in WSN. We have named this novel technique as Recursive Converging Quartiles. It also uses quartiles to summarise a list of readings of any length to just three values. This highly reduces the amount of data to be transmitted to the sink, thus reducing the transmission energy required and at the same time representing the original values accurately.

Another strength of WAPMS is the high quality of results it produces. The collected readings are saved in a database and these can be accessed individually in a table or summarised area-wise in a line graph. The table uses the AQI to provide the results using the associated colours and it also provided the level of health concern for a particular area. The line graph allows the user to view the trend of air pollution for several areas at a time. WAPMS also displays a map of the town of Port Louis, showing the locations of the deployed sensors nodes and the readings collected by each one. Thus, WAPMS is very flexible, very easy and yet very powerful due to its ability to provide highly summarised results as well as fine-grain results at the level of sensors.